\documentclass{article}
\usepackage{latexsym}

\usepackage{hyperref}

\begin{document}
\begin{center}
\begin{title}
\textbf{A DISCONTINUITY OF THE BACKGROUND EXPLAINS THE PIONEER ANOMALY}\\
\end{title}
\vspace{2cm}

Frederic Henry-Couannier\\
CPPM, 163 Avenue De Luminy, Marseille 13009 France.\\
henry@cppm.in2p3.fr\\
\end{center}

\begin{abstract}

The Pioneer anomaly is explained very simply if we assume that somewhere between us and the aircraft, the scale factor 
has undergone a discrete jump from an expansion regime a(t) to a contraction regime 1/a(t).

\end{abstract}
\section{Introduction}
The so called Pioneer anomaly arised as an anomalous drift in time of the radiowave frequency received from both 
Pioneer 10 and 11 as compared to reference clocks on earth \cite{And}. The anomalous drift was found constant over more than 10 years and 
from 40 to 60 AU 
\[
\begin{array}{l}
 \frac{\dot{f}}{f}=(-5.6\pm 0.9) 10^{-18}/s\\ 
 \end{array}
\]
including in the error all systematical source of uncertainties.
Since all possible systematical origins investigated by Anderson and collaborators \cite{And} were excluded or found very unlikely to account for 
the Pioneer anomaly, there is increasing evidence that it is now of fundamental physics origin. Many papers 
have convincingly argued that the answer cannot be found in General Relativity where background effects in the vicinity 
of a local source as is our sun, are completely negligible \cite{Giu}. It was also easily checked that even assuming that expansion takes
 place in the solar system in the same way as it does in empty space far from any matter source, we would at most get $\frac{\dot {f}}{f}= v/c.H0$
 where v stands for the aircraft speed relative to the earth, which is still negligible as compared to the Pioneer effect and has the wrong sign.
More recent studies on the outer planet trajectories in the solar system exclude all attempts so far to explain the effect by 
a universal modification of the gravitational force or the gravitational 
influence of any unknown source in the outer part of the solar system where the anomaly was most significantly detected \cite{Pio1}
\cite{Pio2} .
This also seems to rule out the most usual interpretation of the anomaly in terms of an anomalous acceleration directed 
toward the sun. Instead, various recent attempts seem now to privilege exotic physical effects either affecting the radio wave 
propagation or the clocks drift in time as caused by their supposed interaction with a background field.

\section{A background discontinuity explains the anomaly}

Motivated by the strange coincidence
$ \frac{\dot {f}}{f}=-2\frac{1}{11.10^9\mbox{years}} \sim -2\frac{\dot {a}}{a}\;\equiv -2H_0$
at a roughly 15 percent confidence level coming from both the uncertainty in the directly measured Hubble parameter $H_{0}$ and in the effect, 
 we privilege a background origin for the anomaly in the framework of an hypothetical theory that would treat the background effects 
differently than GR does. Moreover we take serious the fact that the effect after all appears as a drift in time of the Pioneer clocks relative
 to our earth reference clocks. This is enough to suggest that if the anomaly originates from fundamental physics, the new laws must affect 
Pioneer and earth clocks differently. Hence, if it is a background effect, the background must not be the same on earth and beyond 20 AU 
where the anomalies were detected. 
Following this way of thincking it is straigntforward to find how the two different background evolutions must be related in conformal time units
 to account for the effect. We can write down the following metrics on earth and on Pioneer to get the clocks drift in time relative to each other.

On earth:
\begin{equation}
d\tau^2= a^2(t)(dt^2-d\sigma^2) \Rightarrow dt_{earthclock} =\frac{1}{a(t)}d\tau 
\end{equation}

On Pioneer:
\begin{equation}
d\tau^2= a^{-2}(t)(dt^2-d\sigma^2) \Rightarrow  dt_{pioneerclock} =a(t)d\tau 
\end{equation}

The result is a frequency deceleration of Pioneer clocks relative to our earth clocks
$f_{Pioneer} =\frac{1}{a^2(t)}f_{earth} $ yielding the anomalous $\frac{\dot {f}_{Pioneer} }{f_{Pioneer} }=-2H_0$. Therefore, their must be a discrete jump from a(t) to 1/a(t) of the background field between us and Pioneer so that the effect 
could only start to be seen after the crossing of this frontier by the aircraft. Within the error bars the jump 
(see the steep rise up of the effect in \cite{And}) could not have been better evidenced than it was around 15 AU in 1983 by Pioneer 11 .

There is a theory involving discrete symmetries where this kind of jump was predicted, this is the Dark Gravity theory presented 
in much details in \cite{fhc1} and \cite{fhc2}. The theory involves many such discontinuities in motion in the universe and at last can also easily account for 
the Hubble diagram test with Supernovae. Recently a discontinuity in the speed of the solar wind was detected by New Horizons and could well
represent another signature of a background discontinuity as explained in \cite{fhc2} .

\end{document}